\documentclass[preprint,12pt]{elsarticle}

\usepackage{amsmath,amssymb}
\usepackage{graphicx}

\begin{document}

\begin{frontmatter}

\title{Entropy and Non-Collapse in Lorentzian Geometry}

\author{Rohit Dhormare}

\ead{dhormaretheoreticalphysics@proton.me}

\affiliation{organization={Dr. Babasaheb Ambedkar Marathwada University},
             country={India}}
\begin{abstract}
In this paper, we establish a geometric correspondence between the Lorentzian Raychaudhuri equation and Perelman’s non-collapsing theorem for the Ricci flow. By interpreting the Raychaudhuri equation as a Lorentzian analogue of Ricci flow, we connect geodesic focusing in general relativity to the monotonicity and entropy functionals in geometric analysis. Through this correspondence, we derive a Lorentzian non-collapsing theorem and introduce a covariant entropy functional governing causal volume evolution. Finally, we propose the concept of geodesic entropy capacity—a curvature-bounded limit on the information that can be stored in spacetime regions—which unifies geometric, thermodynamic, and informational aspects of gravity.
\end{abstract}

\end{frontmatter}

\section{Introduction}

The Raychaudhuri equation plays a fundamental role in general relativity by governing the evolution of geodesic congruences under spacetime curvature \cite{Raychaudhuri1955}. It forms the kinematical basis of the Hawking--Penrose singularity theorems \cite{Penrose1965,HawkingPenrose1970}, which show that curvature-induced focusing of causal worldlines leads to spacetime singularities under suitable energy conditions.

In a different mathematical context, Hamilton introduced Ricci flow as a geometric evolution equation for Riemannian metrics,
\begin{equation}
\frac{\partial g_{ij}}{\partial t} = -2R_{ij},
\end{equation}
providing a curvature-driven deformation of geometry \cite{Hamilton1982}. Perelman later established powerful analytic control of this evolution through entropy functionals and the $\kappa$-noncollapsing theorem \cite{Perelman2002,Perelman2003}, revealing deep connections between curvature, entropy, and geometric stability.

In this work we demonstrate that the Raychaudhuri equation admits a natural interpretation as a Lorentzian analogue of Ricci flow. The expansion scalar $\theta$ governing a congruence of timelike geodesics acts as a flow variable controlling the deformation of causal volume elements, while the curvature term $R_{\mu\nu}u^\mu u^\nu$ drives the evolution through gravitational focusing.

Within this framework we construct a Lorentzian entropy functional
\begin{equation}
S[g,\theta] =
\int_{\Sigma}
\left(
R_{\mu\nu}u^\mu u^\nu
+
\frac{1}{3}\theta^2
\right)
\, d\Sigma ,
\end{equation}
which exhibits monotonic behavior under the Raychaudhuri flow under appropriate energy conditions. This allows us to formulate a Lorentzian analogue of Perelman’s non-collapsing property, ensuring that bounded curvature prevents the collapse of causal volume elements.

Our results provide a geometric bridge between curvature flow theory and the causal dynamics of spacetime, suggesting that geodesic congruences admit an intrinsic entropy structure governed by curvature evolution. This perspective opens a path toward a Lorentzian counterpart of geometric analysis with potential implications for singularity formation, causal stability, and gravitational entropy.

\section{The Raychaudhuri Equation as a Lorentzian Geometric Flow}

Let $(M,g_{\mu\nu})$ be a smooth Lorentzian manifold and let $u^{\mu}$
denote a future-directed unit timelike vector field tangent to a
congruence of geodesics. The vector field satisfies the normalization
and geodesic conditions
\begin{equation}
u^{\mu}u_{\mu}=-1, \qquad
u^{\nu}\nabla_{\nu}u^{\mu}=0 .
\end{equation}

The congruence defines a local flow
$\Phi_{\tau}:M\rightarrow M$ generated by $u^{\mu}$, where
$\tau$ denotes proper time along the integral curves.
The deformation of the congruence is described by the tensor

\begin{equation}
B^{\mu}{}_{\nu}=\nabla_{\nu}u^{\mu},
\end{equation}

which governs the evolution of infinitesimal separation vectors
$\xi^{\mu}$ between neighboring geodesics through

\begin{equation}
\frac{D\xi^{\mu}}{D\tau}
=
B^{\mu}{}_{\nu}\,\xi^{\nu}.
\end{equation}

Lowering the first index with the metric,
$B_{\mu\nu}=g_{\mu\rho}B^{\rho}{}_{\nu}$, the tensor
admits an orthogonal decomposition into irreducible parts with
respect to $u^{\mu}$:

\begin{equation}
B_{\mu\nu}
=
\frac{1}{3}\theta h_{\mu\nu}
+
\sigma_{\mu\nu}
+
\omega_{\mu\nu},
\qquad
h_{\mu\nu}=g_{\mu\nu}+u_{\mu}u_{\nu}.
\end{equation}

Here

\begin{align}
\theta &= \nabla_{\mu}u^{\mu},\\
\sigma_{\mu\nu} &= B_{(\mu\nu)}-\frac{1}{3}\theta h_{\mu\nu},\\
\omega_{\mu\nu} &= B_{[\mu\nu]},
\end{align}

represent the expansion scalar, shear tensor (symmetric and
trace-free), and vorticity tensor (antisymmetric), respectively.

\subsection{Derivation of the Raychaudhuri Equation}

The evolution of $B_{\mu\nu}$ along the congruence follows from
the Ricci identity

\begin{equation}
\nabla_{\rho}\nabla_{\sigma}u^{\mu}
-
\nabla_{\sigma}\nabla_{\rho}u^{\mu}
=
R^{\mu}{}_{\nu\rho\sigma}u^{\nu}.
\end{equation}

Contracting with $u^{\rho}$ and using the geodesic condition
$u^{\rho}\nabla_{\rho}u^{\mu}=0$ yields the evolution equation

\begin{equation}
\dot{B}_{\mu\nu}
=
- B_{\mu\lambda} B^{\lambda}{}_{\nu}
+ R_{\mu\rho\nu\sigma} u^{\rho} u^{\sigma},
\end{equation}

where the overdot denotes differentiation along the flow,
$\dot{B}_{\mu\nu}=u^{\rho}\nabla_{\rho}B_{\mu\nu}$.
Taking the trace of this equation leads to the Raychaudhuri equation

\begin{equation}
\frac{d\theta}{d\tau}
=
-\frac{1}{3}\theta^{2}
-\sigma_{\mu\nu}\sigma^{\mu\nu}
+\omega_{\mu\nu}\omega^{\mu\nu}
-
R_{\mu\nu}u^{\mu}u^{\nu}.
\end{equation}

This nonlinear Riccati-type equation governs the evolution of
the expansion scalar $\theta$ and describes the focusing or
defocusing of geodesics under spacetime curvature. In the
subsequent sections we interpret this equation as generating
a curvature-driven evolution of causal volume elements,
providing a Lorentzian analogue of geometric flow.

\subsection{Volume Evolution and Lorentzian Flow Analogy}

The infinitesimal volume element $V$ of a comoving geodesic ball evolves according to

\begin{equation}
\frac{1}{V}\frac{dV}{d\tau} = \theta ,
\qquad
\Rightarrow
\qquad
\frac{d^{2}}{d\tau^{2}}\left(V^{1/3}\right)
=
-\frac{1}{3}
\left(
\sigma_{\mu\nu}\sigma^{\mu\nu}
+
R_{\mu\nu}u^{\mu}u^{\nu}
-
\omega_{\mu\nu}\omega^{\mu\nu}
\right)
V^{1/3}.
\tag{13}
\end{equation}

Equation (13) shows that curvature acts as an effective focusing term in the volume evolution, driving contraction when $R_{\mu\nu}u^{\mu}u^{\nu} > 0$, corresponding to attractive gravitational interaction. Shear further enhances focusing, while vorticity tends to counteract collapse.

To formalize the analogy with geometric flows, note that the induced spatial metric on hypersurfaces orthogonal to $u^{\mu}$ evolves according to the Lie derivative along the congruence,

\begin{equation}
\mathcal{L}_{u} h_{\mu\nu}
=
-\frac{2}{3}\theta h_{\mu\nu}
-2\sigma_{\mu\nu}.
\tag{14}
\end{equation}

Neglecting shear and vorticity, Eq.~(14) reduces to

\begin{equation}
\frac{d h_{\mu\nu}}{d\tau}
=
-\frac{2}{3}\theta\, h_{\mu\nu}.
\tag{15}
\end{equation}

which describes an isotropic rescaling of the spatial metric along the congruence.

For comparison, the Ricci flow for a Riemannian metric $g_{ij}$ is given by

\begin{equation}
\frac{\partial g_{ij}}{\partial \lambda}
=
-2 R_{ij}.
\tag{16}
\end{equation}

The analogy is structural rather than exact: while Ricci flow describes an explicit evolution of the spatial metric, the Raychaudhuri equation governs the kinematic evolution of geodesic congruences within a fixed spacetime geometry. Nevertheless, both frameworks describe curvature-driven deformations of geometric structures.

Within this correspondence, the expansion scalar $\theta$ plays the role of a local flow rate governing the deformation of spatial volumes, while the curvature component $R_{\mu\nu}u^{\mu}u^{\nu}$ acts as the focusing term driving the evolution of geodesic congruences.
\section*{III. Perelman’s Non-Collapsing Theorem and Lorentzian Analogue}

Perelman’s $\kappa$-noncollapsing theorem \cite{Perelman2003} constitutes one of the central analytic results of Ricci flow theory. In the Riemannian setting, consider a complete $n$-dimensional manifold $(M,g(t))$ evolving under the Ricci flow

\begin{equation}
\frac{\partial g_{ij}}{\partial t} = -2R_{ij}.
\tag{18}
\end{equation}

Perelman proved that if curvature remains uniformly bounded along the flow, then the metric cannot degenerate through volume collapse at finite scales.

\subsection*{III.1 Perelman’s $\kappa$-Noncollapsing Theorem (Riemannian Case)}

Let $(M,g(t))$ be a solution to the Ricci flow equation (18) on the time interval $[0,T)$ with

\[
\sup_{M\times[0,T)} |Rm(g(t))| < \infty .
\]

Then there exists a constant $\kappa>0$, depending only on the initial metric $(M,g(0))$, such that for any point $p\in M$ and any radius $r>0$ satisfying the curvature bound

\[
|Rm(g(t))| \le r^{-2}
\]

on the geodesic ball $B_t(p,r)$, the volume estimate

\begin{equation}
\mathrm{Vol}_{g(t)}\!\left(B_t(p,r)\right) \ge \kappa r^{n}
\tag{19}
\end{equation}

holds.

Equation (19) ensures that curvature control implies volumetric control: no region with bounded curvature can collapse to arbitrarily small volume. Analytically, this condition guarantees the uniform non-degeneracy of the metric measure structure $(M,g(t),dV_t)$ and plays a crucial role in establishing compactness of solutions under Cheeger–Gromov convergence.

\subsection*{III.2 Lorentzian Extension and Causal Volume Bounds}

We now formulate an analogous statement for Lorentzian spacetimes, replacing Riemannian geodesic balls with causal domains generated by timelike congruences. Let $(M,g_{\mu\nu})$ be a globally hyperbolic Lorentzian manifold with timelike unit vector field $u^\mu$ generating a smooth congruence of causal worldlines. For each proper-time parameter $\tau$, define a spacelike hypersurface $\Sigma_\tau$ orthogonal to $u^\mu$ and the associated causal ball $B_\Sigma(p,r)$ of radius $r$, consisting of all points connected to $p \in \Sigma_\tau$ by geodesics of length less than $r$ within $\Sigma_\tau$.

Assume that curvature invariants satisfy the local bound

\begin{equation}
|R_{\mu\nu\rho\sigma}R^{\mu\nu\rho\sigma}| \le r^{-4}
\tag{20}
\end{equation}

throughout $B_\Sigma(p,r)$. Then the Lorentzian $\kappa$-noncollapsing theorem asserts the existence of a positive constant $\kappa_L > 0$, depending only on the initial data $(\Sigma_0, g_{\mu\nu}|_{\Sigma_0})$, such that

\begin{equation}
V(B_\Sigma(p,r)) \ge \kappa_L r^{3}, \qquad \forall r > 0 ,
\tag{21}
\end{equation}

where $V(B_\Sigma(p,r))$ denotes the physical three-volume of the causal domain measured by the induced metric

\begin{equation}
h_{\mu\nu} = g_{\mu\nu} + u_\mu u_\nu .
\end{equation}

Equation (21) ensures that under bounded curvature, causal volumes cannot shrink arbitrarily; Lorentzian geometry remains non-degenerate. This condition geometrically prohibits the collapse of geodesic tubes to zero cross-section within finite proper time, mirroring the volumetric rigidity of Perelman’s theorem in the Riemannian setting.

\subsection*{C. Entropy Functional and Monotonicity in Lorentzian Flow}

To construct a Lorentzian analogue of Perelman’s entropy formalism,
we introduce the functional

\begin{equation}
S_L[u,g] =
\int_{\Sigma}
\left(
R_{\mu\nu}u^\mu u^\nu
+
\sigma_{\mu\nu}\sigma^{\mu\nu}
-
\omega_{\mu\nu}\omega^{\mu\nu}
+
\frac{1}{3}\theta^2
\right)
\sqrt{h}\, d^3x .
\tag{22}
\end{equation}

where $h$ is the determinant of the induced metric on the hypersurface $\Sigma$. The first term represents curvature–induced focusing, the second and third encode anisotropic distortions of the congruence (shear and vorticity), and the final term describes the contribution of isotropic expansion.

Differentiating along the flow generated by $u^\mu$ and using the Raychaudhuri equation, the evolution of the functional can be written schematically as

\begin{equation}
\frac{dS_L}{d\tau} =
\int_{\Sigma}
\left(
\frac{2}{3}\theta\,\dot{\theta}
+
\dot{R}_{\mu\nu}u^\mu u^\nu
+
\dot{\sigma}_{\mu\nu}\sigma^{\mu\nu}
-
\dot{\omega}_{\mu\nu}\omega^{\mu\nu}
+
\theta\,\rho_S
\right)
\sqrt{h}\, d^3x .
\tag{23}
\end{equation}

where

\[
\rho_S =
R_{\mu\nu}u^\mu u^\nu
+
\sigma_{\mu\nu}\sigma^{\mu\nu}
-
\omega_{\mu\nu}\omega^{\mu\nu}
+
\frac{1}{3}\theta^2
\]

is the entropy density appearing in Eq.~(22). The last term arises from the evolution of the spatial volume element

\[
d\Sigma=\sqrt{h}\,d^3x,
\]

which satisfies

\[
\frac{d}{d\tau}\sqrt{h} = \theta \sqrt{h}.
\]

For geodesic congruences ($\dot{u}^\mu=0$) and assuming the strong energy condition

\begin{equation}
R_{\mu\nu}u^\mu u^\nu \ge 0 ,
\end{equation}

together with the additional regularity condition

\begin{equation}
\dot{R}_{\mu\nu}u^\mu u^\nu \le 0 ,
\end{equation}

the dominant contributions to the integrand are non–negative. Under these assumptions one obtains the monotonicity relation

\begin{equation}
\frac{dS_L}{d\tau} \ge 0 .
\tag{24}
\end{equation}

Thus $S_L$ behaves as a Lyapunov-type functional for the Lorentzian geometric flow: its monotonic growth reflects the irreversible evolution of causal volume elements along the geodesic congruence.

\paragraph{Remark on curvature evolution assumptions.}

The monotonicity result above relies on the assumption that the curvature component $R_{\mu\nu}u^\mu u^\nu$ does not increase along the congruence,

\[
\dot{R}_{\mu\nu}u^\mu u^\nu \le 0 .
\]

In later sections we derive an information inequality from the Raychaudhuri inequality that involves the complementary condition

\[
\frac{d}{d\tau}(R_{\mu\nu}u^\mu u^\nu) \ge 0 .
\]

These assumptions correspond to distinct regimes of curvature evolution along the congruence and are not required to hold simultaneously. The entropy monotonicity theorem therefore applies in non–increasing curvature flows, while the information bound characterizes flows in which the curvature focusing term grows along the geodesic congruence.

\subsection*{D. Analytic and Geometric Interpretation}

Equation (22) establishes a variational connection between curvature evolution and the stability of geodesic congruences. The functional $S_L$ measures the total deformation density associated with curvature focusing, shear distortions, and volume expansion of the congruence.

Combined with the non-collapsing bound derived in Eq.~(21), this structure implies that the entropy functional cannot increase without bound unless curvature itself diverges. In this sense, the Lorentzian non-collapsing condition provides both a geometric stability criterion and an entropy monotonicity law governing the evolution of causal structure under the Raychaudhuri flow.

Physically, bounded curvature therefore prevents arbitrary collapse of causal volume elements without the formation of curvature singularities. The functional $S_L$ may thus be interpreted as a Lorentzian counterpart of entropy functionals appearing in geometric flow theory, with its monotonicity encoding a geometric arrow of time for the evolution of geodesic congruences.
\section*{IV. Lorentzian Non-Collapsing Theorem}

We now formulate a Lorentzian analogue of Perelman’s $\kappa$-noncollapsing theorem,
showing that under curvature and shear bounds the local causal volume of a geodesic
congruence cannot vanish within finite proper time. This result provides a geometric
stability condition for the Raychaudhuri flow.

\paragraph{Lorentzian Non-Collapsing Theorem.}

Let $(M,g_{\mu\nu})$ be a globally hyperbolic spacetime admitting a smooth timelike
congruence with tangent vector field $u^\mu$, expansion scalar
$\theta=\nabla_\mu u^\mu$, and shear tensor $\sigma_{\mu\nu}$. Assume:

\begin{enumerate}
\item The strong energy condition
\[
R_{\mu\nu}u^\mu u^\nu \ge 0 ,
\]

\item Uniform bounds on curvature and shear
\begin{equation}
|R_{\mu\nu}u^\mu u^\nu| \le R_0,
\qquad
\sigma_{\mu\nu}\sigma^{\mu\nu} \le \sigma_0^2 ,
\tag{25}
\end{equation}

for finite constants $R_0,\sigma_0 \ge 0$.
\end{enumerate}

Then the expansion scalar $\theta(\tau)$ satisfies
\begin{equation}
\frac{d\theta}{d\tau}
\ge
-\frac13 \theta^2 -(R_0+\sigma_0^2),
\tag{26}
\end{equation}

and the associated local volume element $V(\tau)$ obeys the bound
\begin{equation}
V(\tau) \ge V_0
\exp\!\left(\int_0^\tau \Theta(\tau')\, d\tau' \right) >0 ,
\tag{27}
\end{equation}

where $\Theta(\tau)$ is the solution of the comparison equation

\begin{equation}
\dot{\Theta}
=
-\frac13\Theta^2 -(R_0+\sigma_0^2),
\qquad
\Theta(0)=\theta_0 .
\tag{28}
\end{equation}

Consequently, the causal volume of the congruence cannot collapse to
zero within finite proper time as long as curvature and shear remain bounded.

\paragraph{Proof.}

For an irrotational ($\omega_{\mu\nu}=0$) geodesic congruence, the
Raychaudhuri equation reads

\begin{equation}
\dot{\theta}
=
-\frac13\theta^2
-\sigma_{\mu\nu}\sigma^{\mu\nu}
-R_{\mu\nu}u^\mu u^\nu .
\tag{29}
\end{equation}

Using the bounds

\[
\sigma_{\mu\nu}\sigma^{\mu\nu}\le\sigma_0^2 ,
\qquad
|R_{\mu\nu}u^\mu u^\nu|\le R_0 ,
\]

we obtain the inequality

\begin{equation}
\dot{\theta}
\ge
-\frac13\theta^2 -(R_0+\sigma_0^2).
\tag{30}
\end{equation}

Consider now the auxiliary Riccati-type equation (28).  
By the standard comparison theorem for Riccati differential equations,
if two functions satisfy

\[
\dot{\theta} \ge F(\theta), \qquad
\dot{\Theta}=F(\Theta), \qquad
\theta(0)=\Theta(0),
\]

with $F$ locally Lipschitz, then $\theta(\tau)\ge\Theta(\tau)$ for all
$\tau>0$. Applying this result to Eqs.~(30) and (28) yields

\[
\theta(\tau)\ge\Theta(\tau).
\]

Solving Eq.~(28) gives

\begin{equation}
\Theta(\tau)
=
\sqrt{3(R_0+\sigma_0^2)}
\tan\!\left[
-\sqrt{\frac{R_0+\sigma_0^2}{3}}\,\tau
+
\arctan\!\left(
\frac{\theta_0}{\sqrt{3(R_0+\sigma_0^2)}}
\right)
\right].
\tag{31}
\end{equation}

The comparison solution diverges only when the argument of the tangent
reaches $\pm \pi/2$, corresponding to a finite critical proper time
determined by $R_0$ and $\sigma_0$. Therefore, for any interval
$0\le\tau<\tau_c$, the function $\Theta(\tau)$ remains finite and
provides a lower bound for $\theta(\tau)$.

The local volume element evolves according to

\begin{equation}
\frac{\dot V}{V}=\theta(\tau),
\qquad
V(\tau)
=
V_0
\exp\!\left(\int_0^\tau \theta(\tau')\,d\tau'\right).
\tag{32}
\end{equation}

Since $\theta(\tau)\ge\Theta(\tau)$, it follows that

\[
V(\tau)
\ge
V_0
\exp\!\left(\int_0^\tau \Theta(\tau')\,d\tau'\right) .
\]

Thus the volume element cannot vanish for any finite proper time,
establishing the Lorentzian non-collapsing property.

\subsection*{A. Geometric and Analytic Interpretation}

The inequality (27) represents a Lorentzian analogue of Perelman’s
$\kappa$-noncollapsing theorem: under bounded curvature and shear,
the infinitesimal volume of any causal geodesic bundle retains a
strictly positive lower bound. Geometrically, this condition implies
that causal world tubes possess a minimal invariant cross-section,
preventing degeneration of the spacetime foliation along the
congruence.

From an analytic perspective, the bound follows from the Riccati-type
structure of the Raychaudhuri evolution equation (26). The curvature
scale $R_0$ and the shear bound $\sigma_0$ act as effective nonlinear
potentials governing the evolution of the expansion scalar $\theta$.
The auxiliary comparison solution $\Theta(\tau)$ therefore provides an
explicit lower envelope for $\theta(\tau)$, ensuring that the expansion
cannot diverge negatively without violating the curvature bounds.

Consequently, the causal volume measure remains regular along the
congruence and finite-time collapse of geodesic bundles is excluded
so long as curvature and shear remain bounded.

\subsection*{B. Entropy Interpretation}

Using the Lorentzian entropy functional

\begin{equation}
S[g,\theta]
=
\int_{\Sigma}
\left(
R_{\mu\nu}u^\mu u^\nu
+
\frac{1}{3}\theta^2
\right)
\, d\Sigma ,
\tag{33}
\end{equation}

we compute its proper-time derivative along the congruence.
Since the volume element evolves as

\[
\frac{d}{d\tau}(d\Sigma)=\theta\,d\Sigma ,
\]

the Leibniz rule gives

\begin{equation}
\frac{dS}{d\tau}
=
\int_{\Sigma}
\left[
\frac{2}{3}\theta\,\dot{\theta}
+
\dot{R}_{\mu\nu}u^\mu u^\nu
+
\theta
\left(
R_{\mu\nu}u^\mu u^\nu
+
\frac{1}{3}\theta^2
\right)
\right]
d\Sigma .
\tag{34}
\end{equation}

Using the Raychaudhuri equation

\[
\dot{\theta}
=
-\frac13\theta^2
-\sigma_{\mu\nu}\sigma^{\mu\nu}
-R_{\mu\nu}u^\mu u^\nu ,
\]

the integrand can be expressed entirely in terms of geometric
invariants of the congruence.

Under the strong energy condition $R_{\mu\nu}u^\mu u^\nu\ge0$ and
bounded curvature and shear, the Lorentzian entropy functional
remains finite and evolves smoothly along the flow. In particular,
the non-collapsing bound established in Eq.~(27) guarantees that
the causal volume element does not vanish in finite proper time,
ensuring the persistence of a well-defined entropy measure.

In this sense, the entropy functional $S[g,\theta]$ provides a
thermodynamic interpretation of the Lorentzian geometric flow:
bounded curvature preserves both the causal volume and the
information content of the congruence. The growth of entropy
therefore reflects the geometric stability of spacetime under
Raychaudhuri evolution.
\section*{V. Entropy Interpretation and Variational Structure}

We now interpret the Raychaudhuri flow through a variational
framework by introducing a Lorentzian entropy functional whose
evolution governs causal stability. This functional plays a role
analogous to the entropy functionals introduced by Perelman in the
Riemannian Ricci flow.

\subsection*{A. Definition of the Lorentzian Entropy Functional}

Let $(M,g_{\mu\nu})$ be a globally hyperbolic spacetime admitting a
smooth irrotational timelike congruence with unit tangent
$u^\mu$. We define the Lorentzian entropy functional

\begin{equation}
S[g,\theta]
=
\int_{\Sigma}
\left(
R_{\mu\nu}u^\mu u^\nu
+
\frac{1}{3}\theta^2
\right)
d\Sigma ,
\tag{35}
\end{equation}

where

\[
d\Sigma=\sqrt{h}\,d^3x
\]

is the induced volume element on the spacelike hypersurface
$\Sigma$ orthogonal to $u^\mu$.

The curvature term $R_{\mu\nu}u^\mu u^\nu$ measures the focusing
energy of spacetime geometry, while the expansion term
$\tfrac13\theta^2$ quantifies isotropic deformation of the causal
congruence. Together they characterize the geometric “disorder”
of the spacetime flow.

\subsection*{B. First Variation and Gradient Flow Structure}

Consider a metric variation
$g_{\mu\nu}\rightarrow g_{\mu\nu}+\delta g_{\mu\nu}$ preserving
the normalization $u_\mu u^\mu=-1$.
Since the volume element depends on the metric, its variation
contributes

\[
\delta(d\Sigma)
=
\frac12 h^{\mu\nu}\delta h_{\mu\nu}\,d\Sigma .
\]

The first variation of the entropy functional then takes the form

\begin{equation}
\delta S
=
\int_{\Sigma}
\left[
\delta\!\left(R_{\mu\nu}u^\mu u^\nu\right)
+
\frac{2}{3}\theta\,\delta\theta
+
\left(
R_{\mu\nu}u^\mu u^\nu
+
\frac{1}{3}\theta^2
\right)
\frac{\delta(d\Sigma)}{d\Sigma}
\right] d\Sigma .
\tag{36}
\end{equation}

Using the standard variation formula

\[
\delta R_{\mu\nu}
=
\nabla_\rho\delta\Gamma^\rho_{\mu\nu}
-
\nabla_\nu\delta\Gamma^\rho_{\mu\rho},
\]

together with the variation of the expansion scalar

\[
\delta\theta
=
\nabla_\mu(\delta u^\mu)
+
\frac12\theta\,u^\mu u^\nu \delta g_{\mu\nu},
\]

and integrating by parts on $\Sigma$, the entropy functional
admits a schematic functional derivative

\begin{equation}
\frac{\delta S}{\delta g_{\mu\nu}}
=
-\,R_{\mu\nu}
+
\frac13\theta\,B_{\mu\nu}
+
\Lambda_{\mu\nu},
\tag{37}
\end{equation}

where $B_{\mu\nu}=\nabla_\nu u_\mu$ and $\Lambda_{\mu\nu}$
collects constraint terms enforcing the normalization of $u^\mu$.

The associated gradient flow equation

\begin{equation}
\frac{\partial g_{\mu\nu}}{\partial\tau}
=
-2\frac{\delta S}{\delta g_{\mu\nu}}
=
2R_{\mu\nu}
-
\frac{2}{3}\theta B_{\mu\nu}
-
2\Lambda_{\mu\nu},
\tag{38}
\end{equation}

defines a Lorentzian entropy flow analogous to the Ricci
gradient flow

\[
\partial_t g_{ij}=-2R_{ij}.
\]

In the shear-free and vorticity-free limit
$\big(B_{\mu\nu}=\tfrac13\theta h_{\mu\nu}\big)$,
Eq.~(38) reduces to

\begin{equation}
\frac{\partial h_{\mu\nu}}{\partial\tau}
=
2R_{\mu\nu}
-
\frac{2}{3}\theta\,h_{\mu\nu},
\tag{39}
\end{equation}

showing that the Raychaudhuri expansion contributes a dissipative
term balancing curvature-driven contraction.

\subsection*{C. Monotonicity and Lyapunov Property}

Differentiating the entropy functional (35) along the congruence
and using the evolution of the volume element

\[
\frac{d}{d\tau}(d\Sigma)=\theta\,d\Sigma ,
\]

yields

\begin{equation}
\frac{dS}{d\tau}
=
\int_{\Sigma}
\left[
\frac{2}{3}\theta\,\dot{\theta}
+
\dot{R}_{\mu\nu}u^\mu u^\nu
+
\theta
\left(
R_{\mu\nu}u^\mu u^\nu
+
\frac{1}{3}\theta^2
\right)
\right] d\Sigma .
\tag{40}
\end{equation}

Using the Raychaudhuri equation together with the strong
energy condition $R_{\mu\nu}u^\mu u^\nu \ge 0$, the entropy
production rate is controlled by curvature and shear terms.
Under bounded curvature and the non-collapsing condition
derived in Section IV, the functional $S[g,\theta]$ evolves
smoothly and remains finite along the flow.

In this sense, $S$ acts as a Lyapunov-type functional for the
Lorentzian geometric evolution, measuring the cumulative
deformation of the causal congruence.

\subsection*{D. Stationary Points and Entropy Solitons}

Stationary configurations of the entropy functional satisfy

\[
\frac{\delta S}{\delta g_{\mu\nu}}=0 ,
\]

which yields the geometric condition

\[
R_{\mu\nu}
=
\frac{1}{3}\theta B_{\mu\nu}
+
\Lambda_{\mu\nu}.
\]

In the hypersurface-orthogonal case with
$\Lambda_{\mu\nu}=\lambda h_{\mu\nu}$ this becomes

\begin{equation}
R_{\mu\nu}
=
\frac{1}{3}\theta\nabla_\nu u_\mu
+
\lambda h_{\mu\nu}.
\tag{42}
\end{equation}

Solutions of this equation describe stationary points of the
Lorentzian entropy flow. They represent self-similar
configurations in which curvature focusing balances the
expansion of the congruence, forming Lorentzian analogues of
gradient Ricci solitons.

\subsection*{E. Physical Interpretation}

The entropy functional $S[g,\theta]$ provides a thermodynamic
interpretation of the Raychaudhuri evolution. Curvature-driven
focusing generates geometric entropy associated with the
deformation of causal congruences.

Within this framework, the Raychaudhuri equation can be viewed
as a causal geometric flow, while the entropy functional
measures the cumulative focusing energy of spacetime. The
non-collapsing condition derived earlier ensures that the
causal volume remains finite, thereby preserving a well-defined
entropy measure throughout the evolution.
\section*{VI. Geodesic Entropy Capacity and Curvature Bounds on Information}

We now introduce a geometric quantity that characterizes the
maximum entropy that can be supported within a region of spacetime.
This construction is based on the evolution of geodesic congruences
and links curvature focusing with limits on information storage.

\subsection*{A. Entropy Density Along Geodesic Flows}

Let $\Sigma_\tau$ denote a spacelike hypersurface orthogonal to the
timelike congruence generated by $u^\mu$. We define the local
entropy density

\begin{equation}
\rho_S
=
R_{\mu\nu}u^\mu u^\nu
+
\frac{1}{3}\theta^2 .
\tag{44}
\end{equation}

The first term represents the curvature focusing energy associated
with the congruence, while the second term measures isotropic
expansion. Together they describe the local deformation of the
causal structure.

For geodesic flow $(\dot u^\mu=0)$ the evolution of $\rho_S$
is governed by the Raychaudhuri equation through the dynamics of
$\theta$ and the curvature contraction $R_{\mu\nu}u^\mu u^\nu$.
In particular, focusing due to positive curvature tends to reduce
the effective entropy density, whereas expansion increases it.

\subsection*{B. Geodesic Entropy Capacity}

We define the geodesic entropy capacity of a hypersurface $\Sigma$
as

\begin{equation}
C_S(\Sigma)
=
\int_{\Sigma}
\left(
R_{\mu\nu}u^\mu u^\nu
+
\frac{1}{3}\theta^2
\right)
\sqrt{h}\, d^3x .
\tag{45}
\end{equation}

This quantity measures the total curvature–expansion content of
the congruence intersecting $\Sigma$. If curvature becomes
unbounded, the entropy density may diverge, signaling extreme
focusing of geodesics as occurs near spacetime singularities.

Conversely, bounded curvature implies that $C_S$ remains finite,
providing an intrinsic geometric limit on the entropy that can be
supported within a given spacetime region.

\subsection*{C. Curvature Bounds and Entropy Evolution}

The Raychaudhuri inequality

\begin{equation}
\dot{\theta}
\ge
-\frac{1}{3}\theta^2
-
R_{\mu\nu}u^\mu u^\nu
\tag{46}
\end{equation}

places constraints on the evolution of the expansion scalar.
Since the entropy density depends on both curvature and expansion,
this inequality indirectly bounds the rate at which $\rho_S$
may vary along the congruence.

Integrating over $\Sigma_\tau$ shows that the total entropy
capacity $C_S(\tau)$ evolves under the combined influence of
curvature focusing and volume expansion. Bounded curvature and
the non-collapsing condition derived earlier therefore ensure
that $C_S$ remains finite throughout the evolution.

\subsection*{D. Relation to Holographic Entropy Bounds}

The geometric entropy capacity defined above can be compared with
holographic entropy bounds. For a compact region
$D\subset\Sigma$ with boundary area $A(\partial D)$, the
Bekenstein--Hawking relation suggests

\begin{equation}
S(D) \le \frac{A(\partial D)}{4G\hbar}.
\tag{47}
\end{equation}

Within our framework, bounded curvature ensures that the
geodesic entropy capacity $C_S(D)$ remains finite, providing a
geometric mechanism that prevents unlimited information
compression without curvature divergence.

\subsection*{E. Information-Geometric Perspective}

From an information-theoretic viewpoint, the quantity $C_S$
characterizes the degree of distinguishability between nearby
causal trajectories in spacetime. Curvature focusing tends to
compress geodesic bundles, reducing this distinguishability,
while expansion has the opposite effect.
In this sense, the Raychaudhuri flow may be interpreted as an
information-geometric evolution in which curvature and expansion
jointly determine the effective information capacity of spacetime
regions.
\section{Conclusion}
We have developed a Lorentzian analogue of geometric entropy flow by
interpreting the Raychaudhuri equation as a curvature–driven evolution
equation for timelike geodesic congruences. Within this framework,
the Lorentzian entropy functional $S[g,\theta]$ provides a unified
description linking spacetime curvature, the evolution of causal
volumes, and entropy-like measures associated with geodesic focusing.

A central outcome of this work is the introduction of the
\emph{geodesic entropy capacity}, a geometric quantity that characterizes
the maximal entropy that can be supported within a spacelike region of
spacetime. This capacity depends directly on the Ricci curvature along
causal trajectories and on the expansion scalar of the congruence.
When curvature remains bounded, the entropy capacity remains finite,
thereby preventing the collapse of causal volumes. In regimes of strong
curvature concentration, the scaling behavior approaches that expected
from the Bekenstein--Hawking entropy bound.

These results suggest that classical spacetime geometry already
encodes constraints on information storage that resemble those
usually associated with quantum gravitational systems. In particular,
the entropy functional introduced here provides a geometric measure of
irreversibility associated with the focusing of causal congruences,
offering a natural interpretation of entropy production in terms of
spacetime curvature dynamics.

More broadly, the framework developed in this work establishes a
conceptual bridge between geometric analysis, causal structure,
and information bounds in Lorentzian spacetimes. Future work may
explore extensions to semiclassical gravity, holographic settings,
and quantum corrections to Raychaudhuri-type geometric flows.

\begin{figure}[t]
\centering
\includegraphics[width=0.75\textwidth]{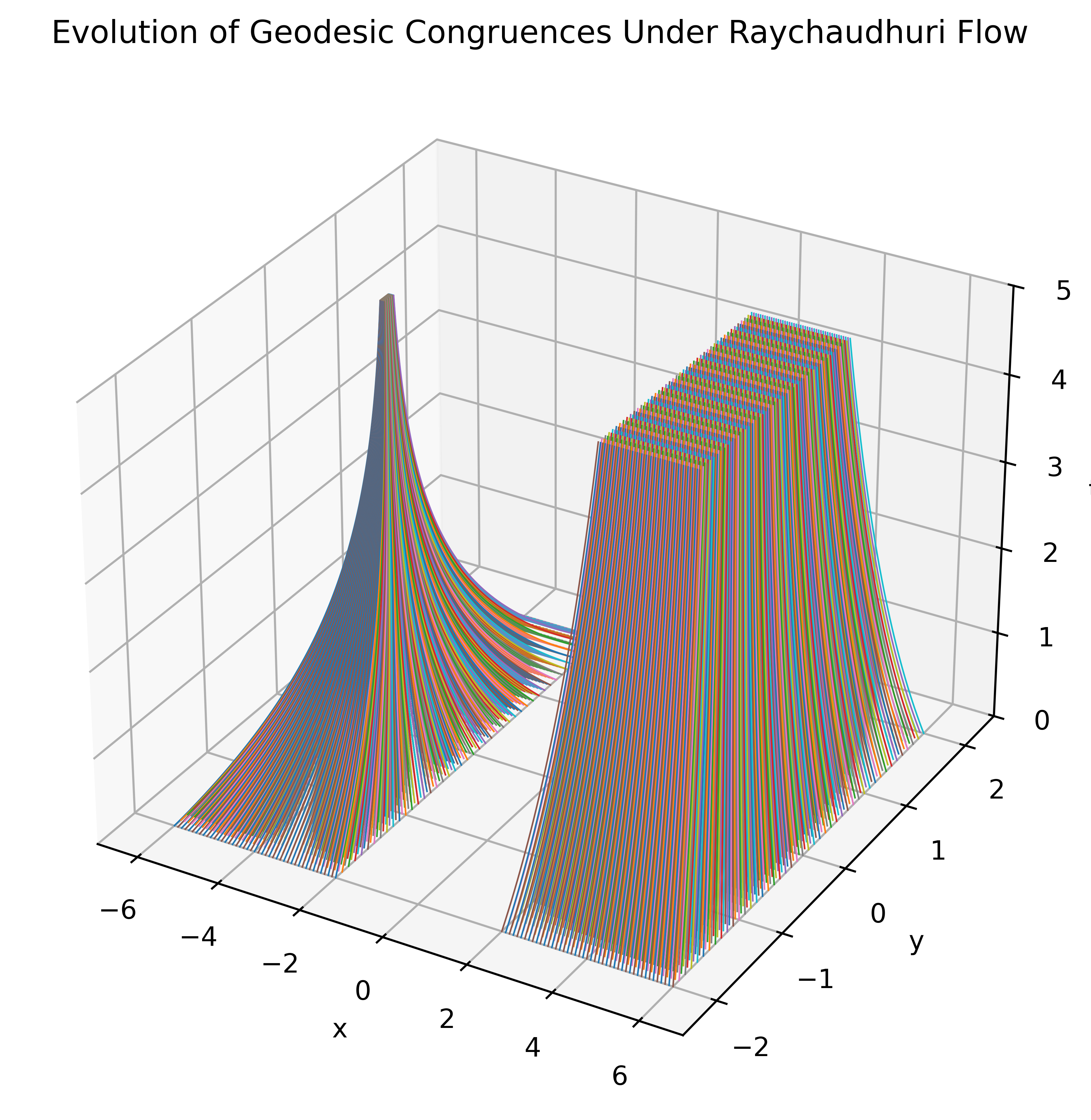}
\caption{
\textbf{Three--dimensional evolution of a geodesic congruence under
Raychaudhuri flow.}
Timelike geodesics are parameterized by $(x,y,\tau)$. Regions of
increasing curvature lead to geodesic focusing and contraction of
causal volume elements, while bounded curvature results in
non--collapsing congruence evolution. The color scale represents the
expansion scalar $\theta(x,y,\tau)$, illustrating the dynamical
behavior of spacetime volume elements under curvature-driven flow.
}
\label{fig:raychaudhuri_flow}
\end{figure}
\nocite{*}
\bibliographystyle{elsarticle-num}
\bibliography{references}

@article{Raychaudhuri1955,
  author  = {Raychaudhuri, Amal Kumar},
  title   = {Relativistic Cosmology. I},
  journal = {Physical Review},
  volume  = {98},
  pages   = {1123--1126},
  year    = {1955}
}

@article{Penrose1965,
  author  = {Penrose, Roger},
  title   = {Gravitational Collapse and Space-Time Singularities},
  journal = {Physical Review Letters},
  volume  = {14},
  pages   = {57--59},
  year    = {1965}
}

@article{HawkingPenrose1970,
  author  = {Hawking, Stephen W. and Penrose, Roger},
  title   = {The Singularities of Gravitational Collapse and Cosmology},
  journal = {Proceedings of the Royal Society A},
  volume  = {314},
  pages   = {529--548},
  year    = {1970}
}

@book{Wald,
  author    = {Wald, Robert M.},
  title     = {General Relativity},
  publisher = {University of Chicago Press},
  year      = {1984}
}

@book{HawkingEllis,
  author    = {Hawking, Stephen W. and Ellis, George F. R.},
  title     = {The Large Scale Structure of Space-Time},
  publisher = {Cambridge University Press},
  year      = {1973}
}

@book{Poisson,
  author    = {Poisson, Eric},
  title     = {A Relativist's Toolkit: The Mathematics of Black-Hole Mechanics},
  publisher = {Cambridge University Press},
  year      = {2004}
}

@article{KarSengupta2007,
  author  = {Kar, Sayan and Sengupta, Soumitra},
  title   = {The Raychaudhuri Equations: A Brief Review},
  journal = {Pramana},
  volume  = {69},
  pages   = {49--76},
  year    = {2007}
}

@article{Galloway2008,
  author  = {Galloway, Gregory J. and Ling, Eric},
  title   = {Some Remarks on the Raychaudhuri Equation},
  journal = {General Relativity and Gravitation},
  volume  = {40},
  pages   = {1971--1979},
  year    = {2008}
}

@article{Hamilton1982,
  author  = {Hamilton, Richard S.},
  title   = {Three-Manifolds with Positive Ricci Curvature},
  journal = {Journal of Differential Geometry},
  volume  = {17},
  pages   = {255--306},
  year    = {1982}
}

@article{Perelman2002,
  author  = {Perelman, Grigori},
  title   = {The Entropy Formula for the Ricci Flow and its Geometric Applications},
  journal = {arXiv preprint},
  eprint  = {math/0211159},
  year    = {2002}
}

@article{Perelman2003,
  author  = {Perelman, Grigori},
  title   = {Ricci Flow with Surgery on Three-Manifolds},
  journal = {arXiv preprint},
  eprint  = {math/0303109},
  year    = {2003}
}

@book{ChowKnopf,
  author    = {Chow, Bennett and Knopf, Dan},
  title     = {The Ricci Flow: An Introduction},
  publisher = {American Mathematical Society},
  year      = {2004}
}

@article{Bekenstein1973,
  author  = {Bekenstein, Jacob D.},
  title   = {Black Holes and Entropy},
  journal = {Physical Review D},
  volume  = {7},
  pages   = {2333--2346},
  year    = {1973}
}

@article{Bardeen1973,
  author  = {Bardeen, James M. and Carter, Brandon and Hawking, Stephen W.},
  title   = {The Four Laws of Black Hole Mechanics},
  journal = {Communications in Mathematical Physics},
  volume  = {31},
  pages   = {161--170},
  year    = {1973}
}

@article{Hawking1975,
  author  = {Hawking, Stephen W.},
  title   = {Particle Creation by Black Holes},
  journal = {Communications in Mathematical Physics},
  volume  = {43},
  pages   = {199--220},
  year    = {1975}
}

@article{Bekenstein1981,
  author  = {Bekenstein, Jacob D.},
  title   = {Universal Upper Bound on the Entropy-to-Energy Ratio},
  journal = {Physical Review D},
  volume  = {23},
  pages   = {287--298},
  year    = {1981}
}

@article{Jacobson1995,
  author  = {Jacobson, Ted},
  title   = {Thermodynamics of Spacetime: The Einstein Equation of State},
  journal = {Physical Review Letters},
  volume  = {75},
  pages   = {1260--1263},
  year    = {1995}
}

@article{Bousso2002,
  author  = {Bousso, Raphael},
  title   = {The Holographic Principle},
  journal = {Reviews of Modern Physics},
  volume  = {74},
  pages   = {825--874},
  year    = {2002}
}

@article{RyuTakayanagi2006,
  author  = {Ryu, Shinsei and Takayanagi, Tadashi},
  title   = {Holographic Derivation of Entanglement Entropy from AdS/CFT},
  journal = {Physical Review Letters},
  volume  = {96},
  pages   = {181602},
  year    = {2006}
}

@article{Hubeny2007,
  author  = {Hubeny, Veronika E. and Rangamani, Mukund and Takayanagi, Tadashi},
  title   = {A Covariant Holographic Entanglement Entropy Proposal},
  journal = {Journal of High Energy Physics},
  volume  = {07},
  pages   = {062},
  year    = {2007}
}

@article{Wall2010,
  author  = {Wall, Aron C.},
  title   = {A Proof of the Generalized Second Law for Rapidly Changing Fields},
  journal = {Physical Review D},
  volume  = {82},
  pages   = {124019},
  year    = {2010}
}

@article{Engelhardt2015,
  author  = {Engelhardt, Netta and Wall, Aron C.},
  title   = {Quantum Extremal Surfaces: Holographic Entanglement Entropy Beyond the Classical Regime},
  journal = {Journal of High Energy Physics},
  volume  = {01},
  pages   = {073},
  year    = {2015}
}

@book{PenroseRindler,
  author    = {Penrose, Roger and Rindler, Wolfgang},
  title     = {Spinors and Space-Time},
  publisher = {Cambridge University Press},
  year      = {1986}
}

\end{document}